\documentclass{article}

\usepackage{arxiv}

\usepackage[dvipdfmx]{graphicx}
\usepackage{url}
\usepackage{listings,jvlisting}
\usepackage{amssymb}
\usepackage{tabularx}
\usepackage{amsmath}
\usepackage{mdframed}
\usepackage{graphicx}
\usepackage{setspace}
\usepackage{here}
\usepackage{authblk}
\usepackage{amsthm}
\usepackage{algorithmic}
\usepackage{algorithm}
\usepackage{arydshln}
\usepackage{longtable}
\usepackage{setspace}
\usepackage[numbers]{natbib}
\usepackage{subcaption}
\usepackage{bbm}
\usepackage{hyperref}
\usepackage{newtxtext}

\newtheorem{theorem}{Theorem}

\newtheorem{example}{Example}

\newtheorem{lemma}{Lemma}
\newtheorem{prop}{Proposition}

\newcommand{\argmax}{\mathop{\rm arg~max}\limits}

\def\bbe{{\text{\boldmath $\beta$}}}

\def\x{{\text{\boldmath $x$}}}

\title{Random-effects meta-analysis via generalized linear mixed models: A Bartlett-corrected approach for few studies}

\author{
    \href{https://orcid.org/0000-0002-1444-0280}{\includegraphics[scale=0.06]{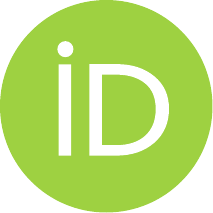}\hspace{1mm}Keisuke~Hanada} \\
    Department of Biostatistics, Faculty of Medicine\\
    Wakayama Medical University\\
    Kimiidera, Wakayama, 641-8509, Japan \\
    \texttt{khanada@wakayama-med.ac.jp} \\
    \and
    \href{https://orcid.org/0000-0001-7609-956X}{\includegraphics[scale=0.06]{orcid.pdf}\hspace{1mm}Tomoyuki~Sugimoto} \\
    Graduate School of Engineering Science \\
    Department of Systems Innovation \\
    Division of Mathematical Science\\
    The University of Osaka\\
    Toyonaka, Osaka 560-8531, Japan\\
    \texttt{sugimoto.tomoyuki.es@osaka-u.ac.jp} \\
}
\date{}

\renewcommand{\shorttitle}{Random-effects meta-analysis via generalized linear mixed models}

\hypersetup{
pdftitle={A template for the arxiv style},
pdfsubject={q-bio.NC, q-bio.QM},
pdfauthor={David S.~Hippocampus, Elias D.~Striatum},
pdfkeywords={First keyword, Second keyword, More},
}

\begin{document}
\maketitle

\begin{abstract}
Random-effects models are central to meta-analysis, yet the between-study variance is often underestimated when the number of studies is small. In such settings, confidence intervals become unduly narrow and fail to attain the nominal coverage probability. Although several small-sample corrections, including the Bartlett correction, have been developed under the normal–normal model, corresponding methodology for generalized linear mixed models (GLMMs) remains limited. This study proposes a unified framework for random-effects meta-analysis within the GLMM that relies exclusively on aggregate data and accommodates outcomes that follow any distribution in the exponential family, including the binomial, Poisson, and gamma distributions. To improve interval estimation with few studies, we develop a profile likelihood method with a simplified Bartlett correction (PLSBC), which refines the chi-squared approximation of the profile likelihood ratio statistic without requiring higher-order derivatives. We show theoretically that the proposed estimators preserve the consistency and asymptotic normality of the maximum likelihood estimators. Simulation studies demonstrate that the PLSBC yields nearly unbiased estimates and maintains nominal coverage across a variety of outcome types. Applications to three published meta-analyses with binomial, Poisson, and gamma outcomes indicate that the proposed approach provides robust and interpretable inference with few studies. The PLSBC therefore offers a practical and broadly applicable framework for random-effects meta-analysis when the number of studies is limited.
\end{abstract}

\keywords{Bartlett correction, few studies, generalized linear model, meta-analysis, random-effects model.}

\section{Introduction}
When a meta-analysis includes only a small number of studies, the variance of the estimated overall treatment effect is often underestimated, and confidence and prediction intervals may have coverage probabilities below the nominal level \citep{li1994bias, viechtbauer2005bias}. Similar deficiencies have been reported for generalized linear mixed models (GLMMs), where empirical investigations have documented undercoverage in meta-analyses based on odds ratios or risk ratios when the number of studies is small \citep{seide2019likelihood}. This issue is particularly critical in areas where the number of available studies is inherently limited, such as research on rare diseases. The primary objective of this study is to develop a meta-analytic inference procedure within the GLMM framework that yields valid interval estimation even when the number of studies is small.

\begin{figure}[ht]
\centering
\includegraphics[width=1.0\linewidth]{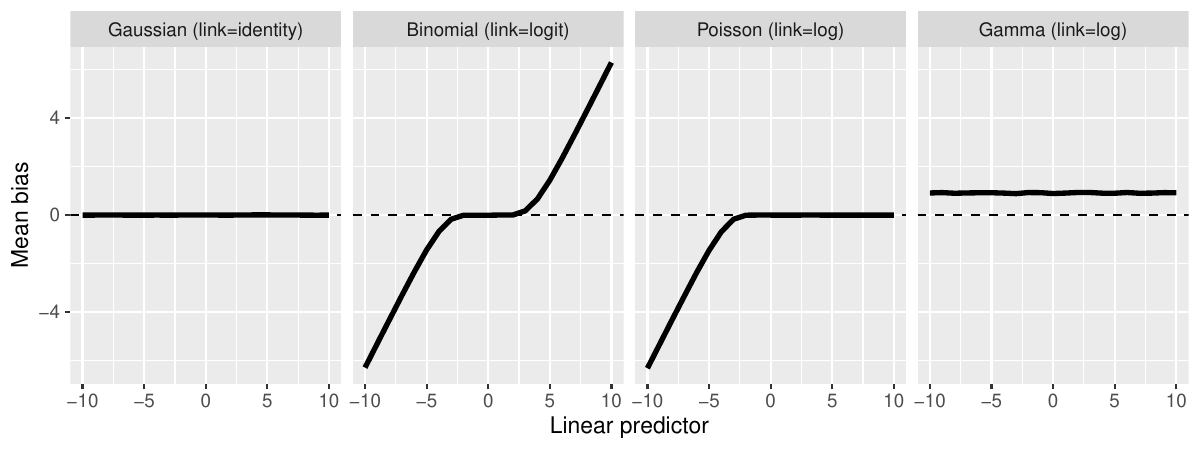}
\caption{Mean bias over 10{,}000 iterations when fitting normal, binomial, Poisson, and gamma models to 20 observations. The binomial and Poisson models incorporate small-sample bias correction. The gamma model is specified with dispersion parameter 30.}
\label{fig-bias-example}
\end{figure}

Conventional random-effects meta-analysis, most commonly implemented through the method of \citet{dersimonian1986meta}, is simple and widely used but implicitly relies on three within-study normality assumptions: (A1) the estimated effect in each study is unbiased; (A2) the within-study variance is known; and (A3) the estimated effect follows a normal distribution \citep{jackson2018should}. In practice, these assumptions are frequently violated. A well-known example is meta-analysis with the log-odds ratio as the outcome, where the estimator is biased and the magnitude of the bias depends on the within-study sample size \citep{bhaumik2012meta}. As illustrated in Figure~\ref{fig-bias-example}, similar biases arise for count data typically modeled by the Poisson distribution and for length-of-stay data typically modeled by the gamma distribution. This bias persists even as the number of studies increases and leads to undercoverage of the associated confidence intervals.

To mitigate bias induced by the normal approximation, several authors have proposed GLMMs that model event counts directly through binomial distributions or their extensions \citep{stijnen2010random, bhaumik2012meta, jackson2018comparison}. GLMMs have also been increasingly used in meta-analyses to estimate effects across diverse types of response variables \citep{platt1999generalized, turner2000multilevel, hamza2008binomial, chu2012bivariate, bohning2015meta, bakbergenuly2018meta, lin2020meta}. These methods avoid assumptions (A1)–(A3) by modeling the outcome distribution directly, but they are typically tailored to specific data types, such as binary or count outcomes. There is thus a need for a more general framework that accommodates outcomes following any distribution in the exponential family.

A small number of studies represents another major cause of undercoverage in confidence intervals. To improve the coverage accuracy of confidence and prediction intervals under assumptions (A1)–(A3), several small-sample corrections have been proposed \citep{michael_exact_2019, nagashima2019prediction, hanada2023inference}. However, these methods rely on assumptions (A1)–(A3), and their performance deteriorates when assumption (A1) is violated, which frequently occurs when GLMMs are used for non-normal outcomes. Although most studies may be large, the presence of even a study with a small within-study sample size can induce bias in the estimator of the treatment effect. This bias does not vanish even as the number of studies increases, and leads to substantial undercoverage of the corresponding confidence intervals (see Web Appendix~A for details). 
This limitation underscores the need for developing small-sample corrections within the GLMM framework. \citet{noma2011confidence} was the first to demonstrate the usefulness of the Barttlet correction in meta-analysis under assumptions (A1)–(A3), although its theoretical foundations had already been established for a broader class of models, including GLMMs \citep{bartlett1953approximate}. Its application to GLMMs, however, remains limited because it requires third- and fourth-order derivatives of the profile likelihood, which depend on the link function and model structure. To overcome this computational complexity, bootstrap-based estimation of the Bartlett correction term has been proposed and applied to network meta-analysis \citep{rocke1989bootstrap, noma2018Bartlett}, but this approach requires repeated likelihood optimization for each bootstrap sample and is computationally demanding. A computationally efficient and generalizable Bartlett correction applicable to the GLMM framework is therefore needed.

Previous work has mainly focused on small-sample corrections under the normal–normal model, which is a random-effects meta-analysis model that assumes normality for both study-level estimates and random effects, whereas corresponding developments for GLMM-based meta-analysis remain limited. This study addresses this gap by establishing a unified GLMM framework that provides valid inference with few studies for outcomes following exponential-family distributions. The proposed method generalizes the conventional normal–normal model to accommodate outcomes following any exponential-family distribution and applies a simplified version of the Bartlett correction originally developed under the normality assumption \citep{noma2011confidence}. The method requires only a single optimization of the profile likelihood, avoids higher-order derivative calculations, and provides accurate and computationally efficient interval estimation that maintains nominal coverage even with few studies. In contrast to existing GLMM-based corrections that are restricted to specific models such as the binomial distribution \citep{mathes2018comparison, seide2019likelihood, beisemann2020comparison}, the proposed method offers a unified framework applicable to exponential-family models, including binomial, Poisson, and gamma outcomes.

The proposed approach has been implemented in the R package \texttt{metaGLMM} (\url{https://github.com/keisuke-hanada/metaGLMM}). The remainder of this paper is organized as follows. Section~\ref{sec-GLMM-meta} introduces the GLMM framework for meta-analysis under the exponential-family setting. Section~\ref{sec-bartlett} describes the proposed profile likelihood method with a simplified Bartlett correction (PLSBC), which is particularly effective in improving coverage accuracy for few studies. Section~\ref{sec-numerical-results} evaluates the finite-sample performance of the proposed method through simulation studies, and Section~\ref{sec-re-meta} presents applications to real meta-analyses. Section~\ref{sec-discussion} concludes with a summary of the main findings and a discussion of directions for future research. Technical details on bias under the normal–normal model, numerical procedures, and proofs are provided in the Web Appendix.

\section{Random-effects meta-analysis under the generalized linear mixed model}
\label{sec-GLMM-meta}
Consider a meta-analysis of $K$ studies, where $n_k$ denotes the number of subjects in study $k$. Between-study heterogeneity is modeled by random effects $V_1, \dots, V_K$, each assumed to follow a probability density function $f_V(\cdot; \tau^2)$ with mean zero and variance $\tau^2$. We denote the realizations of these random effects by $v_1, \dots, v_K$, which are unobserved, but this notion will be convinient in what follows.

\subsection{Model under the exponential-family assumption}
To clarify the underlying data-generating mechanism, we first consider the setting in which individual participant data are available. For study $k$, let $Y_{ki}$ denotes the outcome for subject $(i = 1, \dots, n_k)$. The random variables $Y_{k1}, \dots, Y_{kn_k}$ are assumed to be independently and identically distributed within each study. For each subject $i$ in study $k$, we specify the outcome model
\begin{align}
\label{eq-individual-model}
\mu_{k} &= E[Y_{ki} \mid V_k = v_k], \quad \theta_k = g(\mu_k) = \x_k^T \bbe + v_k,
\end{align}
where $\mu_k$ denotes the conditional mean of $Y_{ki}$ given the study-specific random effect $v_k$, and $\x_k = (x_{k1}, \dots, x_{kp})^T$ is a $p$-dimensional covariate vector for study $k$. The link function $g(\cdot)$ connects the study mean $\mu_k$ with the linear predictor $\theta_k$ and enables inference through a regression framework. The coefficient vector $\bbe = (\beta_1, \dots, \beta_p)^T$ represents the effects of the study-level covariates. Although the canonical link function is often adopted for its natural compatibility with the exponential family, any one-to-one link function is theoretically valid. For example, the identity link $g(\mu_k) = \mu_k$ is canonical for normal outcomes, and the logit link $g(\mu_k) = \mathrm{logit}(\mu_k)$ is canonical for binomial outcomes. The inverse link $g(\mu_k) = 1 / \mu_k$ is canonical for gamma outcomes, although the log link $g(\mu_k) = \log \mu_k$ is often preferred in practice for numerical stability.

Standard analyses based on individual participant data use covariates at the individual level. In contrast, meta-analysis typically relies on study-level aggregates. We therefore assume that a common covariate vector $\x_k$ is available for each study $k$. Examples of such study-level covariates include the average age, sex ratio, and study year, as illustrated in Section~\ref{sec-re-meta}. Given the study-level covariates $\x_k$, the individual observations $(y_{k1}, \dots, y_{kn_k})$ are independently and identically sampled within each study. At the meta-analysis stage, the available information from study $k$ usually consists of $\{n_k, \bar{y}_k, \x_k, \hat{\theta}_k, \hat{\sigma}_k^2\}$, where $n_k$ is the sample size, $\bar{y}_k = n_k^{-1} \sum_{i=1}^{n_k} y_{ki}$ is the sample mean, $\x_k$ denotes the study-level covariates, $\hat{\theta}_k = g(\bar{y}_k)$ is the estimate of $\theta_k$, and $\hat{\sigma}_k^2$ is the estimated variance of $\hat{\theta}_k$.

When individual participant data are available, the likelihood function for study $k$ is
\begin{align}
\label{eq-general-model}
L_k(\bbe, \tau^2)
= \int_{\Omega} \left\{ \prod_{i=1}^{n_k} f_Y(y_{ki}; \bbe, v_k) \right\} f_V(v_k; \tau^2) \, dv_k,
\end{align}
where $f_Y(\cdot)$ denotes the density or probability mass function of $Y$, and $\Omega$ is the support of the random effect $V_k$. In general, evaluation of the likelihood in \eqref{eq-general-model} requires access to individual-level data $y_{ki}$. In aggregate-data meta-analysis, individual-level observations are unavailable, and the inverse-variance weighted estimator is commonly used \citep{dersimonian1986meta}. However, when the outcomes involve nonlinear transformations, such as the log-odds ratio, this normal-approximation estimator ``does not eliminate study-specific bias'', even as the number of studies increases (see Web Appendix~A for a detailed discussion).

To address this limitation, we consider a model that directly reflects the assumed outcome distribution in each study. Specifically, we assume that the outcome follows a GLMM. Under this assumption, the distribution of $Y_{ki}$ belongs to the exponential family and can be written as
\begin{align}
\label{eq-exponential-family}
f_Y(y_{ki}; \bbe, v_k)
= \exp\left( \frac{y_{ki}(\x_k^T \bbe + v_k) - b(\x_k^T \bbe + v_k)}{a(\varphi_k)} + c(y_{ki}, \varphi_k) \right),
\end{align}
where $\varphi_k$ is a dispersion parameter, $a(\varphi_k)$ and $b(\x_k^T \bbe + v_k)$ are known distribution-specific functions, and $c(y_{ki}, \varphi_k)$ is a normalizing term that ensures correct probabilistic scaling.

\subsection{Estimation using aggregate data under the exponential family}
Having specified the individual-level model, we now turn to estimation based on study-level aggregate data. In most meta-analyses, individual participant data are not available, and only study-level summaries such as means and standard errors are reported. It is therefore necessary to reformulate the likelihood in terms of these aggregate quantities while preserving the properties of the original GLMM. This formulation links the study-specific mean $\bar{y}_k$ to the marginal expectation of $Y_{ki}$ under the GLMM and thus provides a coherent connection between the individual-level and aggregate-level representations in meta-analysis.

Although the individual outcomes $y_{ki}$ are unobserved, the study-level mean $\bar{y}_k$ is typically available. Because, in exponential-family models, the sum $\sum_{i=1}^{n_k} Y_{ki}$ (and hence $\bar{y}_k$) is a sufficient statistic for the natural parameter conditional on $V_k = v_k$, and because the random-effect density $f_V$ does not depend on $Y_{ki}$ or $\x_k$, the following result holds.

\begin{lemma}
\label{th-exp-sufficient}
Suppose that $Y_{ki}$ follows the exponential-family distribution in \eqref{eq-exponential-family} and that $V_k$ has density $f_V(\cdot; \tau^2)$. Then, conditional on $(\x_1, \dots, \x_K)$, the vector $(\bar{y}_1, \dots, \bar{y}_K)$ is a sufficient statistic for $(\bbe, \tau^2)$. The likelihood function for study $k$ can be expressed as
\begin{align}
\label{eq-ad-ll}
L_k(\bbe, \tau^2; \bar{y}_k, \x_k, \varphi_k)
&= \left( \prod_{i=1}^{n_k} e^{c(Y_{ki}, \varphi_k)} \right)
\int_{\Omega} \exp\left[ \frac{n_k \left\{ \bar{y}_{k} (\x_k^T \bbe + v_k) - b(\x_k^T \bbe + v_k) \right\}}{a(\varphi_k)} \right] f_V(v_k; \tau^2) \, dv_k.
\end{align}
\end{lemma}

The proof of Lemma~\ref{th-exp-sufficient} is provided in Web Appendix~C. Lemma~\ref{th-exp-sufficient} shows that when $Y_{ki}$ follows an exponential-family distribution, the parameters $(\bbe, \tau^2)$ can be estimated solely from the study-level means $\bar{y}_k$, even in the absence of individual-level observations. This property follows from the fact that in exponential-family models, the sample mean $\bar{y}_k$ (or equivalently, the sum $\sum_{i=1}^{n_k} y_{ki}$) is a sufficient statistic for the natural parameter. Consequently, all information about $(\bbe, \tau^2)$ contained in the individual observations is preserved in $\bar{y}_k$. Details of the numerical calculation of \eqref{eq-ad-ll} are provided in Web Appendix~B.

The parameters $(\bbe, \tau^2)$ are estimated by maximizing the log-likelihood
\begin{align}
\label{eq-est-ll}
(\hat{\bbe}, \hat{\tau}^2) = \argmax_{(\bbe, \tau^2)} \log L(\bbe, \tau^2),
\end{align}
where $L(\bbe, \tau^2) = \prod_{k=1}^K L_k(\bbe, \tau^2)$. Although this optimization does not admit a closed-form solution, it can be carried out numerically using only the study-level data $\{n_k, \bar{y}_k, \x_k, \hat{\sigma}_k^2\}_{k=1}^K$ without access to individual-level data. Because the estimator in \eqref{eq-est-ll} is derived from the same likelihood as that based on the full individual-level data, it inherits the consistency and asymptotic normality of the maximum likelihood estimator under standard regularity conditions \citep{hoadley1971asymptotic, philippou1975asymptotic}. This approach therefore enables reliable inference even when individual studies have small sample sizes or when effect measures involve nonlinear transformations, provided that the number of studies or the total sample size increases.

\section{Profile likelihood interval estimation with simplified Bartlett correction}
\label{sec-bartlett}

\subsection{Profile likelihood inference under GLMM}
When the number of studies is small, the between-study variance $\tau^2$ tends to be underestimated, which leads to too narrow confidence intervals for $\bbe$. To improve inference with few studies, we adopt the profile likelihood approach. Let $\beta_{\ell}$ denote the parameter of interest, and let $\bbe(\beta_{\ell})$ be $\bbe$ in which $\beta_\ell$ is held fixed while $\beta_1, \dots, \beta_{\ell-1}, \beta_{\ell}, \beta_{\ell+1}, \dots, \beta_p$ are allowed to vary. The profile likelihood ratio statistic is asymptotically chi-squared:
\begin{align*}
T(\beta_{\ell}^0)
&= -2 \left\{ \log L(\hat{\bbe}(\beta_{\ell}^0), \tilde{\tau}^2(\beta_\ell^0)) - \log L(\hat{\bbe}, \hat{\tau}^2) \right\}
\xrightarrow{d} \chi^2_1 \quad (K \to \infty),
\end{align*}
where $\chi^2_1$ denotes the chi-squared distribution with one degree of freedom, $\xrightarrow{d}$ denotes convergence in distribution, and the constrained maximum likelihood (CML) estimators under the restriction $\beta_{\ell} = \beta_{\ell}^0$ are defined as
\begin{align*}
(\hat{\bbe}(\beta_{\ell}^0), \tilde{\tau}^2(\beta_\ell^0))
&= \argmax_{(\beta_1,\dots, \beta_{\ell-1}, \beta_{\ell+1}, \dots, \beta_p, \tau^2)} \log L(\bbe(\beta_{\ell}^0), \tau^2).
\end{align*}
Under this asymptotic approximation, the profile likelihood yields confidence intervals with nominal coverage when the number of studies is sufficiently large. However, when $K$ is small (for example, $K = 5$), the nominal level is not guaranteed. To address this small-sample issue, the Bartlett correction was proposed \citep{bartlett1953approximate}. In the context of linear mixed models, it was introduced by \citet{zucker2000improved} and later applied to meta-analysis under the normal–normal model by \citet{noma2011confidence}. Under the normal–normal model, the Bartlett correction reduces the approximation error of the chi-squared distribution of $T(\beta_{\ell}^0)$ from $O(K^{-1})$ to $O(K^{-2})$ \citep{barndorff1988level}. Although the underlying theory is applicable to a broad class of likelihood-based models, its use in GLMMs remains limited because evaluation of the correction term requires third- and fourth-order derivatives of the profile likelihood.

To apply the Bartlett correction in the GLMM, we first formulate its expression within the GLMM framework. 
Let $T_{BC}(\beta_{\ell}^0)$ denote the Bartlett-corrected statistic:
\begin{align*}
T_{BC}(\beta_{\ell}^0) = \frac{T(\beta_{\ell}^0)}{1 + 2 C_{BC}(\beta_{\ell}^0)},
\end{align*}
where the correction term $C_{BC}(\beta_{\ell}^0)$ is given by Equation~(4) in \citet{lawley1956general}:
\begin{align*}
C_{BC}(\beta_{\ell}^0)
= \frac{1}{2K}
\left\{
l_2^{-2} \left( \frac{1}{4} l_4 - l_{31} + l_{22} \right)
- l_2^{-3} \left( \frac{5}{12} l_3^2 - 2 l_3 l_{21} + 2 l_{21}^2 \right)
\right\},
\end{align*}
where
\begin{align*}
l_r &= E\left[\frac{\partial^r l}{\partial \beta_{\ell}^r} \right], \quad
l_{rs} = \frac{\partial^s l_r}{\partial \beta_{\ell}^s}, \quad
l(\beta_{\ell}) = \sum_{k=1}^K \log L_k(\hat{\bbe}(\beta_{\ell}), \tilde{\tau}^2), \quad r = 1,2,3,4, \quad s = 1,2,
\end{align*}
and all expectations and derivatives are evaluated at $\beta_{\ell} = \beta_{\ell}^0$.

\subsection{Simplified Bartlett correction}
Direct evaluation of $C_{BC}(\beta_{\ell})$ requires third- and fourth-order derivatives of the log-likelihood, which are analytically intractable when $L(\bbe, \tau^2)$ involves numerical integration under the GLMM. To circumvent this difficulty, we approximate the exact Bartlett correction $C_{BC}(\beta_{\ell})$ by a simpler term $C_{SBC}(\beta_{\ell})$, originally derived under the normal–normal model \citep{noma2011confidence}. The term $C_{SBC}(\beta_{\ell})$ retains the leading $O(K^{-2})$ bias adjustment of $C_{BC}(\beta_{\ell})$ while avoiding higher-order derivatives. Specifically,
\begin{align*}
C_{SBC}(\beta_{\ell}^0)
= \frac{ \sum_{k=1}^K (\sigma_k^2 + \tilde{\tau}^2)^{-3} }
{ \left\{ \sum_{k=1}^K (\sigma_k^2 + \tilde{\tau}^2)^{-1} \right\}
\left\{ \sum_{k=1}^K (\sigma_k^2 + \tilde{\tau}^2)^{-2} \right\} },
\end{align*}
where $\sigma_k^2$ denotes the within-study variance of $\hat{\theta}_k$, and $\tilde{\tau}^2$ is the CML estimate under $\beta_{\ell} = \beta_{\ell}^0$. The corresponding likelihood ratio statistic with the simplified Bartlett correction is
\begin{align*}
T_{SBC}(\beta_{\ell}^0) = \frac{T(\beta_{\ell}^0)}{1 + 2 C_{SBC}(\beta_{\ell}^0)}.
\end{align*}
Because $C_{SBC}(\beta_{\ell}^0) > 0$, this correction produces more conservative intervals than those based directly on $T(\beta_{\ell}^0)$ and improves the asymptotic approximation with respect to the number of studies. Intuitively, the Bartlett correction adjusts the curvature of the profile likelihood surface so that the resulting statistic more closely matches its finite-sample distribution. The simplified version retains this improvement without the need for third- or fourth-order derivatives.

For the number of studies $K$ and within-study sample sizes $n_1, \dots, n_K$, the following result holds.

\begin{theorem}
\label{th-plbc}
Suppose that $Y_{ki}$ follows the exponential-family distribution in \eqref{eq-exponential-family} and that $V_1, \dots, V_K \sim N(0, \tau^2)$. Then, for any $\beta_{\ell} = \beta_{\ell}^0$,
\begin{align*}
T_{SBC}(\beta_{\ell}^0)
&= T_{BC}(\beta_{\ell}^0) + O_p(n^{-1/2})
= \chi_1^2 + O_p(n^{-1/2} + K^{-2}),
\end{align*}
where $n_k = n a_k$ with $\sum_{k=1}^K a_k = 1$ and $a_k > 0$.
\end{theorem}

The proof of Theorem~\ref{th-plbc} is provided in Web Appendix~C. Here, $O_p(n^{-1/2})$ denotes a stochastic term bounded in probability at rate $n^{-1/2}$, in the sense that for any $\varepsilon > 0$ there exists $M > 0$ such that $P(|O_p(n^{-1/2})| > M n^{-1/2}) < \varepsilon$ for sufficiently large $n$. The notation $O_p(K^{-2})$ is defined analogously. Thus, the approximation improves as both the within-study sample sizes and the number of studies increase.

On the basis of Theorem~\ref{th-plbc}, the $100(1-\alpha)\%$ confidence interval for $\beta_{\ell}$ is defined by
\begin{align*}
T_{SBC}(\beta_{\ell}) \le q_{\chi^2_1, \alpha},
\end{align*}
where $q_{\chi^2_1, \alpha}$ denotes the upper $100\alpha\%$ quantile of the chi-squared distribution with one degree of freedom.

The simplified Bartlett correction has several desirable properties. First, under the normal–normal model, $C_{BC}(\beta_{\ell}^0) = C_{SBC}(\beta_{\ell}^0)$, and hence $T_{SBC}(\beta_{\ell}^0) = \chi^2_1 + O_p(K^{-2})$. The chi-squared approximation therefore improves as $K$ increases, regardless of the within-study sample sizes. Second, since $C_{SBC}(\beta_{\ell}^0) > 0$, the corrected statistic always satisfies $T_{SBC}(\beta_{\ell}^0) < T(\beta_{\ell}^0)$ and produces wider, more conservative confidence intervals. Third, because $C_{SBC}(\beta_{\ell}^0) = O(K^{-1})$, we have $T_{SBC}(\beta_{\ell}^0) \to T(\beta_{\ell}^0)$ in probability as $K \to \infty$, and the asymptotic behavior of the standard profile likelihood is preserved. These properties indicate that the simplified Bartlett correction yields confidence intervals that remain valid and moderately conservative when the number of studies is small, while converging to the standard profile-likelihood interval as $K$ increases. From a practical perspective, this correction mitigates the risk of excessively narrow intervals and reduces the likelihood of overconfident conclusions in clinical research.

\begin{figure}[ht]
\centering
\includegraphics[width=1.0\linewidth]{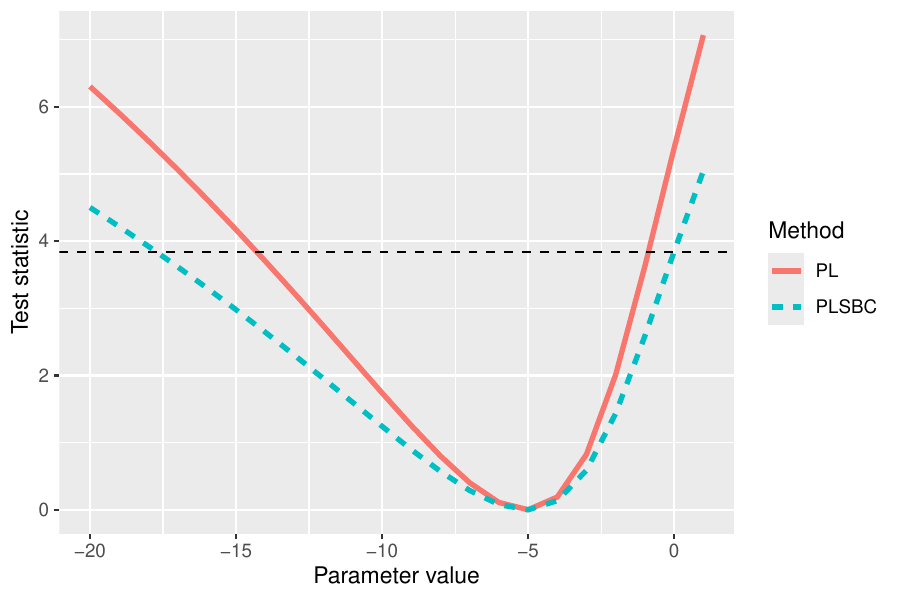}
\caption{Comparison of the profile likelihood (PL) and simplified Bartlett-corrected (PLSBC) test statistics for $\beta_1$ in the reanalysis of \citet{chu2012bivariate} (Section~\ref{sec-re-meta}). The dashed horizontal line indicates the chi-squared threshold $q_{\chi^2_1, 0.05}$.}
\label{fig-bc-effect}
\end{figure}

\section{Simulation study}
\label{sec-numerical-results}
Theoretical results in Theorem~\ref{th-plbc} indicate that the simplified Bartlett correction improves the finite-sample approximation of the profile likelihood ratio. To assess its performance under realistic conditions, we conducted simulation studies for several outcome types.

\subsection{Simulation design}
We conducted Monte Carlo simulations to compare the proposed profile likelihood method and the simplified Bartlett correction with existing meta-analytic methods across different outcome distributions and numbers of studies. Performance was evaluated in terms of bias, coverage probability, and the length of the 95\% confidence interval. The following five methods were considered: (i) the DerSimonian–Laird estimator under the normal–normal model (nDL); (ii) the Bartlett-corrected conservative interval estimation proposed by \citet{noma2011confidence} (nPLBC); (iii) the exact inference under the normal–normal model developed by \citet{michael_exact_2019} (nMI); (iv) the likelihood ratio-based confidence interval under the GLMM (gPL); and (v) the simplified Bartlett-corrected interval under the GLMM (gPLSBC).

Individual participant data were generated from four distributions for $f_Y$: normal, binomial, Poisson, and gamma. The data-generating model assumed normally distributed random effects $V_1, \dots, V_K \sim N(0, \tau^2)$ and was specified as
\begin{align*}
\mu_k = E[Y_{ki} \mid V_k = v_k], \quad \theta_k = g(\mu_k) = \theta_0 + v_k.
\end{align*}
The definitions of the link function $g(\mu_k)$, the transformed parameter $\theta_k$, the dispersion function $a(\varphi_k)$, and the cumulant function $b(\theta_k)$ for each model, as well as further implementation details, are provided in Web Appendix~D. We set $\theta_0 = -2$ and first considered scenarios with fixed between-study variance $\tau^2 = 1$ and number of studies $K = 5, 8, 10, 30, 50$. We then fixed $K = 5$ and varied the between-study variance as $\tau^2 = 0.5, 1.0, 2.0$. The study-specific sample size $n_k$ was generated as the integer part of a random draw from the uniform distribution $U(15, 150)$. All simulation results were based on 10{,}000 replications.

\subsection{Simulation results}
\begin{figure}[H]
\centering
\includegraphics[width=1.0\linewidth]{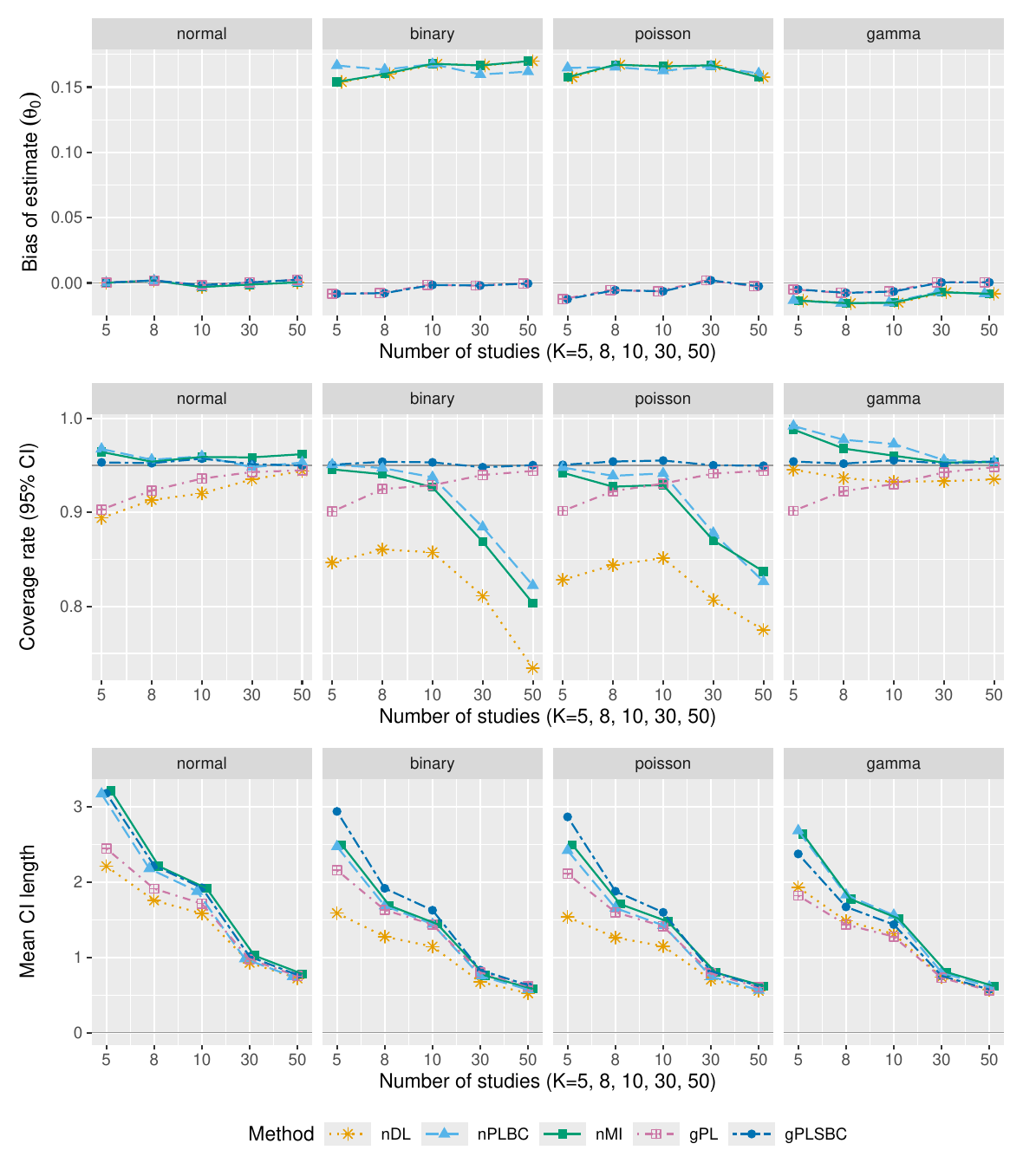}
\caption{
Simulation results for the between-study variance $\tau^2 = 1$ and the number of studies $K = 5, 8, 10, 30, 50$. Each panel presents the bias of the estimated overall effect $\theta_0$ (top), the coverage probability of the 95\% confidence interval (middle), and the average interval length (bottom) across methods. Methods: nDL (DerSimonian–Laird estimator, normal–normal model); nPLBC (profile likelihood with Bartlett correction, normal–normal model); nMI (exact inference, normal–normal model); gPL (profile likelihood, GLMM); gPLSBC (profile likelihood with simplified Bartlett correction, GLMM).}
\label{fig-sim2}
\end{figure}

\begin{figure}[H]
\centering
\includegraphics[width=1.0\linewidth]{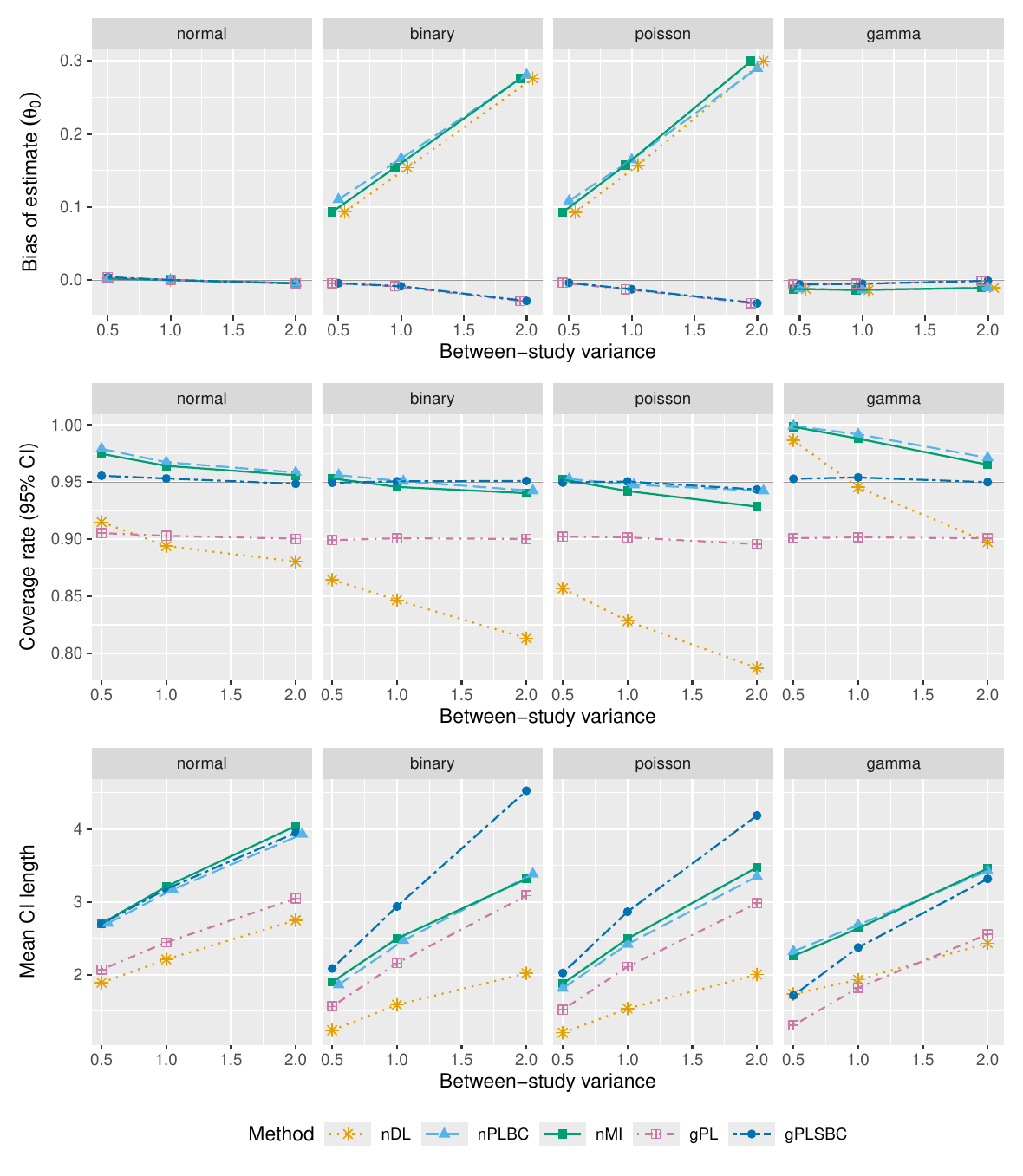}
\caption{
Simulation results for the number of studies $K = 5$ and the between-study variance $\tau^2 = 0.5, 1.0, 2.0$. Each panel presents the bias of the estimated overall effect $\theta_0$ (top), the coverage probability of the 95\% confidence interval (middle), and the average interval length (bottom) across methods. Methods: nDL (DerSimonian–Laird estimator, normal–normal model); nPLBC (profile likelihood with Bartlett correction, normal–normal model); nMI (exact inference, normal–normal model); gPL (profile likelihood, GLMM); gPLSBC (profile likelihood with simplified Bartlett correction, GLMM).}
\label{fig-sim1}
\end{figure}

Under the normal–normal model, all methods are expected to perform well because the model assumptions are exactly satisfied. This setting serves as a benchmark for evaluating the effect of model misspecification under the binomial, Poisson, and gamma models. The results for the scenario with fixed between-study variance $\tau^2 = 1$ are presented in Figure~\ref{fig-sim2}, and those for the scenario with fixed number of studies $K = 5$ are shown in Figure~\ref{fig-sim1}.

\paragraph{Bias}
Under the normal–normal model, all methods (nDL, nMI, nPLBC, gPL, and gPLSBC) produced unbiased estimates of the overall effect, as expected. For non-normal outcomes, methods based on the normal–normal model (nDL, nMI, and nPLBC) exhibited substantial bias under the binomial and Poisson models and moderate bias under the gamma model. This bias results from the nonlinearity of the transformed outcomes, as illustrated in Figure~\ref{fig-bias-example}. Because assumption (A1) is violated for non-normal outcomes, the bias does not vanish as the number of studies increases and becomes more pronounced as $\tau^2$ increases (see Web Appendix~A for details). The GLMM-based methods (gPL and gPLSBC) showed slight bias when the number of studies was small, but the bias diminished as $K$ increased and became negligible once $K \ge 10$.

\paragraph{Coverage probability}
For the 95\% confidence intervals, all methods maintained nominal coverage under the normal–normal model, which shows that the proposed approach performs comparably to existing methods when model assumptions hold. For non-normal outcomes, nDL consistently yielded coverage below the nominal level under the binomial and Poisson models because of bias and also failed to maintain nominal coverage under the gamma model when the between-study variance was large. As $K$ increased, coverage deteriorated further in the binomial and Poisson settings. Although nMI and nPLBC maintained nominal coverage when $K = 5$, their coverage probabilities decreased as $K$ increased. Under the gamma model, nPLBC maintained nominal coverage for all combinations of $K$ and $\tau^2$ but tended to be overly conservative when $K$ was small. The gPL method achieved nominal coverage when $K$ was sufficiently large but underperformed in scenarios with few studies. In contrast, gPLSBC consistently maintained the nominal coverage level across all outcome distributions, values of $\tau^2$, and numbers of studies, and was therefore the most reliable method among those compared.

\paragraph{Interval length}
With respect to the length of the 95\% confidence intervals, nDL produced the shortest intervals in most settings, followed by gPL. Methods that incorporated the Bartlett correction (nPLBC and gPLSBC) or provided conservative exact inference (nMI) produced longer intervals. 
For the binomial and Poisson models, gPLSBC yielded intervals that were slightly longer than those from nPLBC and nMI, whereas under the gamma model, nPLBC produced longer intervals than both gPLSBC and nMI.

\paragraph{Summary}
Overall, under the normal–normal model, all methods yielded unbiased estimates and confidence intervals with nominal coverage, which confirms that the proposed method behaves comparably to existing approaches when model assumptions are satisfied. For non-normal outcomes, the GLMM-based methods (gPL and gPLSBC) provided nearly unbiased estimation even in scenarios where nDL, nPLBC, and nMI exhibited bias. The nMI method provides an exact inference under the normal–normal model, and the original study showed robustness to the choice of random-effects distribution \citep{michael_exact_2019}: nMI maintained nominal coverage even when the random effects followed non-normal distributions such as centered chi-squared or uniform. Nevertheless, the assumption of normality for the outcome distribution, particularly assumption (A1), remains crucial. When within-study sample sizes are finite, the unbiasedness of study-level estimators fails, and confidence intervals based on the normal model show undercoverage. Even for methods such as nMI that are robust to the random-effects distribution, violation of (A1) arises because outcomes from the GLMM are approximated by a normal distribution, and this approximation causes the coverage probability to fall below the nominal level. These findings emphasize that specification of an appropriate outcome model under the GLMM framework is essential for valid interval estimation. In particular, gPLSBC achieved accurate and stable interval estimation that maintained nominal coverage regardless of the outcome distribution, the magnitude of between-study variance, or the number of studies. The gPLSBC method therefore reduces overconfident inference even in the presence of small studies or non-normally distributed outcomes.

Although the proposed gPLSBC method has not been investigated previously, its performance is consistent with earlier reports that GLMM-based meta-analytic approaches outperform normal–normal models in the presence of non-normality \citep{stijnen2010random, hamza2008binomial, jackson2018comparison, bakbergenuly2018meta}. The benefit of the PLSBC is most evident when both the number of studies and the within-study sample sizes are small, which are precisely the conditions under which conventional methods tend to underestimate uncertainty.

\section{Reanalysis of three real-data meta-analyses}
\label{sec-re-meta}
This section presents three meta-analyses in which the outcomes follow binomial, Poisson, and gamma distributions, respectively. Three published meta-analyses corresponding to these outcome types were reanalyzed. For each dataset, point estimates and 95\% confidence intervals from the normal–normal (NN) model (DL, PLBC, MI) and from the GLMM framework (PL, PLSBC) were compared, with particular emphasis on bias-sensitive effect measures and performance with few studies.

\subsection{Data analysis: binomial outcome}
We first reanalyzed the COVID-19 physical distancing data reported by \citet{chu2020physical}. The original study conducted a systematic review and meta-analysis to evaluate the effectiveness of maintaining a distance greater than one meter in preventing SARS-CoV-2 transmission. Among the included studies, seven nonrandomized studies were analyzed in the COVID-19 subgroup and are used here for illustration. Each study compared infection risk between two groups: within one meter ($z = 0$) and beyond one meter ($z = 1$). Let $(r_{kz}, n_{kz})$ denote the number of infections and the total number of participants in arm $z$ of study $k$. The observed infection proportion in arm $z$ of study $k$ is $\bar{y}_{kz} = r_{kz}/n_{kz}$.

The group-specific outcomes were modeled using a binomial GLMM with logit link
\begin{align*}
\theta_{kz} = \log \frac{p_{kz}}{1 - p_{kz}} = \beta_0 + \beta_1 z + V_{kz},
\end{align*}
where $p_{kz}$ denotes the infection probability and $V_{kz}$ represents the random effect. Following \citet{stijnen2010random} and \citet{bhaumik2012meta}, several specifications for $V_{kz}$ have been proposed, such as $V_{kz} = zV_k$ or $V_{kz} = (2z - 1)V_k / 2$. In the present analysis, we assumed $V_{kz} \sim N(0, \tau^2)$ independently across all groups and studies, which allows heterogeneity at both the study and group levels. This specification provides sufficient flexibility in settings with few studies and yields a reasonable approximation even when a shared study-level random effect is present. The corresponding log-likelihood function is
\begin{align*}
\log L(\bbe, \tau^2)
\propto \sum_{k=1}^K \sum_{z=0}^1
\log \int_{\Omega}
\exp \left\{ n_{kz}\bar{y}_{kz}\theta_{kz} - n_{kz}\log(1 + e^{\theta_{kz}}) \right\}
f_V(v_{kz}; \tau^2) \, dv_{kz}.
\end{align*}

In the original article, the DerSimonian–Laird method under the normal–normal model was used. Here, we reanalyzed the data using both NN-based methods (DL, PLBC, MI) and GLMM-based methods (PL, PLSBC). For the NN-based analyses, log-odds ratios with bias correction were computed using the \texttt{metafor} package in R \citep{viechtbauer2010conducting}. Wald-type confidence intervals with bias correction are known to show poor coverage in the presence of zero-event cells \citep{fagerland2015recommended}. Consequently, NN-based methods that rely on bias-corrected log-odds ratios tend to be biased toward the null and to produce overly narrow intervals. For example, in Study~2, the bias-corrected estimate was $-4.905$ (95\% confidence interval [CI]: $[-9.051, -0.760]$), whereas the conditional maximum likelihood estimate based on Fisher’s exact method was $-\infty$ (95\% CI: $[-\infty, -0.580]$).

\begin{figure}[ht]
\centering
\includegraphics[width=1.0\linewidth]{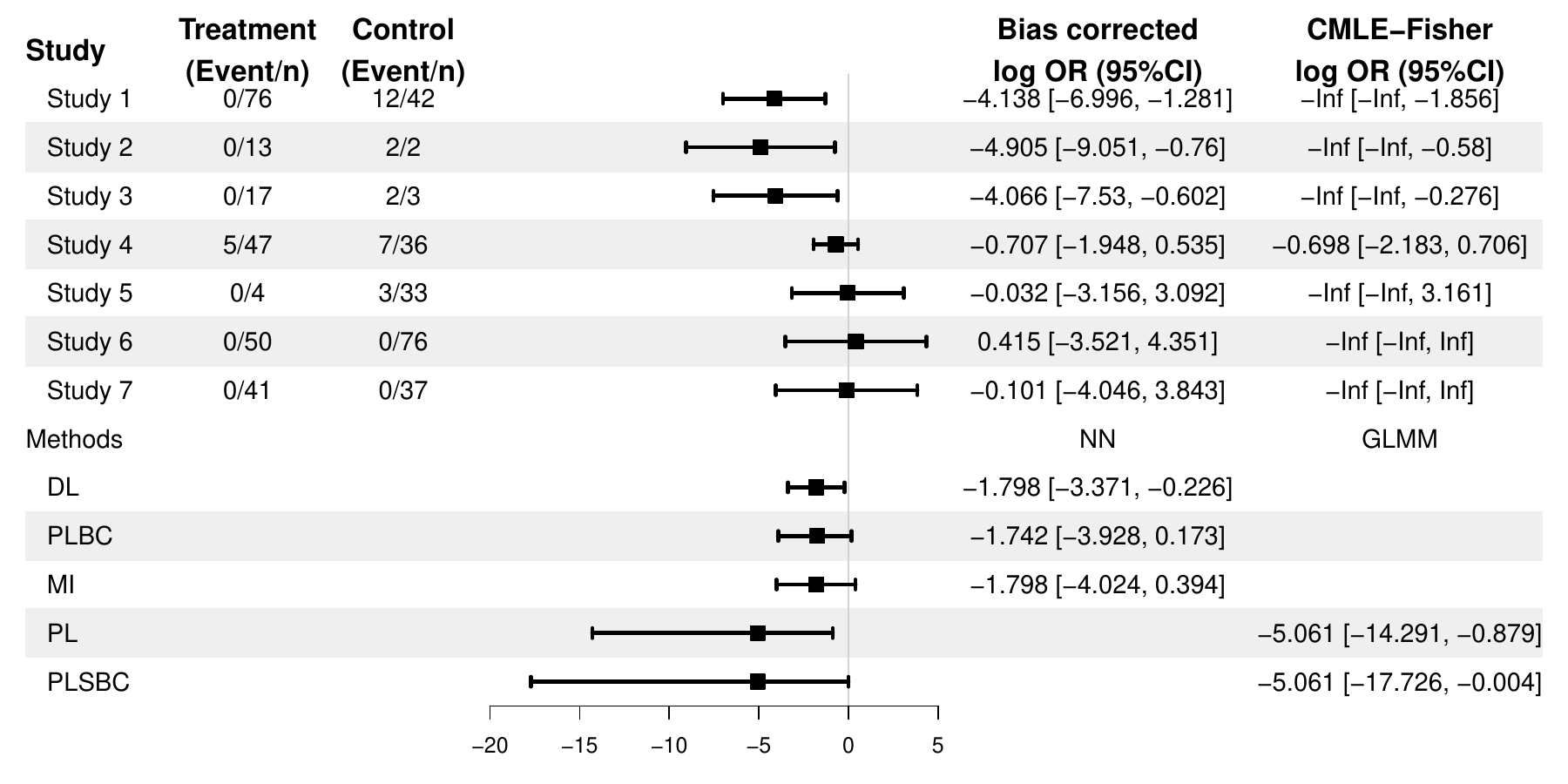}
\caption{
Reanalysis of \citet{chu2020physical}. DL: DerSimonian–Laird method; PL: profile likelihood; PLBC: profile likelihood with Bartlett correction; MI: exact method by \citet{michael_exact_2019}; PLSBC: profile likelihood with simplified Bartlett correction; NN: normal–normal model; GLMM: generalized linear mixed model.}
\label{fig-rda1}
\end{figure}

Figure~\ref{fig-rda1} presents a forest plot that compares the original and reanalyzed results. The GLMM-based estimates were $\mathrm{PL} = -5.061$ \ $[-14.291, -0.879]$ and $\mathrm{PLSBC} = -5.061$ \ $[-17.726, -0.004]$, which more appropriately reflected the between-group differences while maintaining satisfactory coverage. These findings indicate that for binomial outcomes with zero-event cells, the GLMM-based PLSBC method provides more conservative and robust interval estimation than conventional normal–normal approaches and thereby helps to avoid overconfident conclusions in meta-analyses with few studies.

This example illustrates that the PLSBC method yields stable inference even in the presence of zero-event cells and provides a practical approach for meta-analyses with few studies and binary outcomes.

\subsection{Data analysis: Poisson outcome}
Takayasu arteritis (TAK) is a rare large-vessel vasculitis with a marked female predominance. \citet{rutter2021systematic} conducted a systematic review and meta-analysis of eleven population-based or center-based cohort studies published between 2009 and 2018 to estimate the global incidence of TAK. A sex-stratified subgroup analysis (Figure~2 in \citealp{rutter2021systematic}) reported a substantially higher incidence among women ($2.00$ per million person-years; 95\% CI: $[1.29, 3.11]$) than among men ($0.28$ per million person-years; 95\% CI: $[0.14, 0.55]$), consistent with this predominance. Because this sex difference is likely attributable to hormonal, genetic, and immunological mechanisms, sex-stratified reporting is standard in TAK epidemiology.

In our reanalysis, we evaluated temporal trends and sex-specific differences in incidence by including sex, publication year, and their interaction as covariates. Let $y_k$ denote the total number of events and $t_k$ the total person-time in study $k$. Under a Poisson outcome model, $y_k \sim \mathrm{Poisson}(\mu_k)$, and the logarithm of the incidence rate is modeled as
\begin{align*}
\log \mu_k
= \log t_k + \beta_0\ (\mathrm{Intercept}) + \beta_1\ (\mathrm{Sex}) + \beta_2\ (\mathrm{Year}) + \beta_3\ (\mathrm{Sex} \times \mathrm{Year}) + V_k,
\end{align*}
where $\log t_k$ is an offset and $V_k \sim N(0, \tau^2)$ represents study-specific heterogeneity. This specification defines a Poisson GLMM for rate-type data, with inference based on aggregate counts rather than individual observations. The associated log-likelihood function is
\begin{align*}
\log L(\bbe, \tau^2)
\propto \sum_{k=1}^K
\log \int_{\Omega}
\exp \left\{ y_k \log \mu_k - \mu_k \right\}
f_V(v_k; \tau^2) \, dv_k,
\end{align*}
where $\mu_k = t_k \exp(\x_k^T \bbe + v_k)$.

\begin{figure}[H]
\centering
\includegraphics[width=1.0\linewidth]{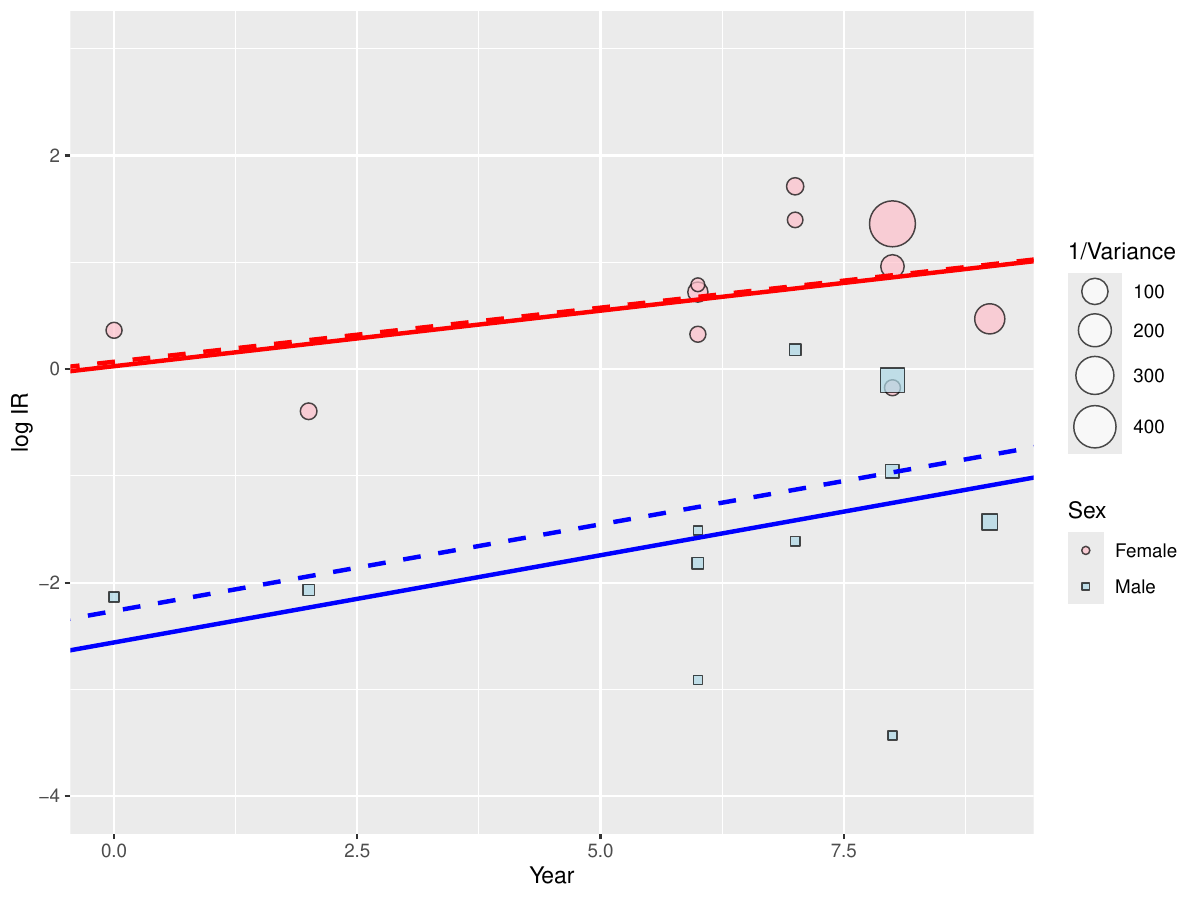}
\caption{
Fitted regression lines for the log incidence rate by year and sex in the reanalysis of \citet{rutter2021systematic}. Red lines correspond to females and blue lines to males. Solid lines indicate the normal–normal model and dashed lines indicate the GLMM. Point sizes are proportional to the inverse of the estimated variance ($1/\widehat{\mathrm{Var}}$), which reflects the relative precision of each study.}
\label{fig-rda-rutter2021}
\end{figure}

We compared the results from the NN model (DL) with those from the GLMM-based methods (PL and PLSBC). Because PLBC and MI are applicable only to univariate models, DL was used for the NN analysis. Table~\ref{tab-rda-rutter2021} summarizes the estimated regression coefficients and their 95\% confidence intervals, and Figure~\ref{fig-rda-rutter2021} shows the fitted regression lines by sex and year. The fitted lines for females were nearly identical between the NN and GLMM analyses, whereas a modest discrepancy appeared for males. Under the GLMM, the estimated coefficients were slightly smaller, with a reduction of approximately 0.2 in the coefficient for sex. Overall, the NN and GLMM analyses led to consistent substantive conclusions, and the confidence intervals from the GLMM with PLSBC were slightly wider, which reflects more conservative uncertainty quantification. These results suggest that in this example the normal–normal model provides reasonable inference, whereas the GLMM framework serves as a useful sensitivity analysis to assess robustness to distributional misspecification. When event rates are moderate and normal approximations are adequate, the GLMM and normal–normal approaches yield comparable results, and the proposed framework remains appropriate under conditions where normal approximations are adequate.

\begin{table}[H]
\centering
\begin{tabular}{|l|cc|ccc|}
\hline
& \multicolumn{2}{c|}{NN} & \multicolumn{3}{c|}{GLMM} \\
\hline
& Estimate & 95\% CI & Estimate & 95\% CI (PL) & 95\% CI (PLSBC) \\
\hline
$\beta_0$ & 0.069 & $[-0.941, 1.079]$ & 0.027 & $[-1.057, 1.106]$ & $[-1.125, 1.174]$ \\
$\beta_1$ & -2.334 & $[-4.151, -0.517]$ & -2.588 & $[-4.628, -0.748]$ & $[-4.753, -0.645]$ \\
$\beta_2$ & 0.101 & $[-0.049, 0.252]$ & 0.104 & $[-0.058, 0.266]$ & $[-0.068, 0.276]$ \\
$\beta_3$ & 0.061 & $[-0.199, 0.321]$ & 0.059 & $[-0.212, 0.343]$ & $[-0.228, 0.361]$ \\
\hline
\end{tabular}
\caption{Estimated regression coefficients and 95\% confidence intervals from the reanalysis of \citet{rutter2021systematic}. NN: normal–normal model; GLMM: generalized linear mixed model; PL: profile likelihood; PLSBC: profile likelihood with simplified Bartlett correction.}
\label{tab-rda-rutter2021}
\end{table}

\subsection{Data analysis: gamma outcome}
\citet{long2020clinical} conducted a random-effects meta-analysis of intensive care unit (ICU) length-of-stay (LoS) across seven randomized controlled trials that compared surgical with conservative management for patients with pneumothorax. The reported pooled mean difference was $-5.72$ days in favor of surgical treatment. Because ICU LoS is strictly positive and typically right-skewed, the normal–normal model may describe the mean–variance relationship and tail behavior inadequately. We therefore reanalyzed these data using a gamma-distributed GLMM with log link.

Let $Y_{ki}$ denote the ICU LoS for subject $i$ in study $k$. Assume $Y_{ki} \sim \mathrm{Gamma}(\gamma_{1k}, \gamma_{2k})$ with shape parameter $\gamma_{1k}$ and scale parameter $\gamma_{2k}$. The study-level sufficient statistics are the sample mean $\bar{y}_k$ and sample variance $s_k^2$. Define
\begin{align*}
\hat{\theta}_k = \log \bar{y}_k, \qquad \hat{\sigma}_k^2 = \frac{s_k^2}{\bar{y}_k^2},
\end{align*}
which represent the estimated mean on the log scale and its estimated variance. At the meta-analysis level, the log mean is modeled as
\begin{align*}
\theta_k = \log \mu_k = \beta_0\ (\mathrm{Intercept}) + \beta_1\ (\mathrm{Treatment}) + V_k, \qquad V_k \sim N(0,\tau^2),
\end{align*}
where $V_k$ represents between-study heterogeneity. The corresponding log-likelihood function, expressed in terms of the study-level sufficient statistics $(\bar{y}_k, s_k^2)$, can be written as
\begin{align*}
\log L(\bbe, \tau^2)
\propto \sum_{k=1}^K
\log \int_{\Omega}
\exp \left\{ \frac{n_k \bar{y}_k \theta_k + n_k \log(\theta_k)}{\hat{\sigma}_k^2} \right\}
f_V(v_k; \tau^2) \, dv_k.
\end{align*}
We estimated the treatment effect $\beta_1$ using the profile likelihood (PL) and the simplified Bartlett correction (PLSBC) under the GLMM and compared these estimates with those from the normal–normal model (DL, PLBC, MI). Table~\ref{tab-rda-long2020} summarizes the study-level data and the estimated treatment effects.

Under the normal–normal model, the estimated coefficient for treatment was $-0.473$, which corresponds to a ratio $\exp(-0.473) \approx 0.623$ and indicates that surgical management reduced ICU LoS to approximately 62\% of that under conservative treatment. The GLMM yielded a similar point estimate $\exp(-0.431) \approx 0.650$, corresponding to approximately 65\% of the control LoS, but produced slightly wider confidence intervals, particularly under PLSBC. The PLSBC-based interval included zero, which suggests that although the estimated treatment effect remains clinically meaningful, the associated uncertainty becomes larger when small-sample bias in each study is properly accounted for.

\begin{table}[H]
\centering
\begin{tabular}{|l|ccc|ccc|}
\hline
& \multicolumn{3}{c|}{Treatment} & \multicolumn{3}{c|}{Control} \\
Study & $n$ & Mean & SD & $n$ & Mean & SD \\
\hline
Study 1 & 20 & 9.6 & 0.7 & 20 & 14.6 & 2.2 \\
Study 2 & 25 & 9.9 & 8.3 & 25 & 10.9 & 11.6 \\
Study 3 & 23 & 13.8 & 4.2 & 23 & 23.3 & 18.7 \\
Study 4 & 18 & 16.5 & 7.4 & 19 & 26.8 & 13.2 \\
Study 5 & 75 & 8.2 & 4.3 & 89 & 14.6 & 3.2 \\
\hline\hline
Method & Estimate & \multicolumn{2}{c|}{95\% CI} & & & \\
\hline
DL    & -0.473 & $[-0.578,$ & $-0.367]$ & & & \\
PLBC  & -0.469 & $[-0.634,$ & $-0.294]$ & & & \\
MI    & -0.473 & $[-0.661,$ & $-0.281]$ & & & \\
PL    & -0.431 & $[-0.682,$ & $-0.042]$ & & & \\
PLSBC & -0.431 & $[-0.896,$ & $0.034]$  & & & \\
\hline
\end{tabular}
\caption{
Reanalysis of \citet{long2020clinical}. DL: DerSimonian–Laird method under the normal–normal model; PLBC: profile likelihood with Bartlett correction under the normal–normal model; MI: exact method by \citet{michael_exact_2019}; PL: profile likelihood under the GLMM; PLSBC: profile likelihood with simplified Bartlett correction under the GLMM.}
\label{tab-rda-long2020}
\end{table}

As indicated by Theorem~\ref{th-plbc}, the simplified Bartlett correction yields more conservative intervals when either the number of studies or the within-study sample sizes is small. Therefore, if a treatment effect remains statistically significant after application of the PLSBC, the result provides solid evidence of a genuine effect even in settings with few studies. This analysis illustrates the advantage of the PLSBC for skewed and heteroscedastic outcomes, where conventional normal–normal methods may underestimate uncertainty. The additional conservativeness does not weaken inference; instead, it ensures that any detected treatment effect remains credible under variability due to few studies and small within-study sample sizes.

\section{Conclusion}
\label{sec-discussion}
This study developed a unified inferential method for random-effects meta-analysis within the GLMM framework. The proposed approach accommodates a broad class of outcome types in the exponential family and remains valid even when the number of studies is small. A simplified Bartlett correction (PLSBC) was introduced to improve interval estimation without higher-order derivatives. Simulation studies and reanalyses of published meta-analyses showed that the PLSBC yields nearly unbiased estimates and confidence intervals with accurate coverage across binomial, Poisson, and gamma outcomes.

Compared with conventional normal–normal models, the GLMM-based framework provides valid inference by modeling the outcome distribution directly and achieves reliable estimation even when both the number of studies and the within-study sample sizes are small. Although the method applies to outcome models within the exponential family, extensions to settings outside this class, such as proportional hazard models, remain important directions for future research. Overall, the proposed GLMM-based meta-analytic framework offers theoretically justified and practically useful inference that combines flexibility with robustness, even with few studies and small within-study sample sizes, across a wide range of applications involving exponential-family outcomes.

\section*{Acknowledgements}
This work was supported by JSPS KAKENHI (Grant numbers JP24K23862 and JP24K14853).
\vspace*{-8pt}

\bibliography{main.bib} 
\bibliographystyle{unsrtnat}

\appendix
\section*{Appendix}

This Appendix complements the main text. Appendix~A examines bias and coverage distortion under the normal--normal model, clarifying why conventional random-effects meta-analyses can remain biased even as the number of studies increases. Appendix~B describes the numerical evaluation of the GLMM likelihood using Monte Carlo and quasi--Monte Carlo integration. Appendix~C provides formal proofs of the main propositions and theorems within a unified asymptotic framework. Appendix~D summarizes the data-generating mechanisms and parameter settings for each outcome distribution used in the simulation studies.

\section{Bias and coverage distortion under the normality assumption}
\label{sec-nn-meta}

In meta-analyses conducted under the normality assumption, the following three conditions are typically required \citep{jackson2018comparison}:

\begin{enumerate}
    \item[(A1)] $\hat{\theta}_k$ is an unbiased estimator: $E[\hat{\theta}_k \mid V_k = v_k] = \theta_k = \x_k^T \bbe + v_k$.
    \item[(A2)] The within-study variance $\sigma_k^2$ is known.
    \item[(A3)] $\hat{\theta}_k$ follows a normal distribution conditional on $V_k = v_k$: $\hat{\theta}_k \mid V_k = v_k \sim N(\theta_k, \sigma_k^2)$.
\end{enumerate}

Although these assumptions may hold asymptotically, they often fail when individual studies have small sample sizes, leading to biased or invalid inference. In such settings, theoretical guarantees depend primarily on the number of studies or the total sample size rather than on large within-study samples.

For clarity, consider a univariate meta-analysis with $p = 1$, $\x_k = 1$, and $\bbe = \theta^*$. Under assumptions (A1)--(A3) and $V_k \sim N(0, \tau^2)$, equation~(1) in the main text can be expressed as the marginal model for $\hat{\theta}_k$:
\begin{align}
\label{eq-nn-model}
    \hat{\theta}_k &\sim N(\theta^*, \sigma_k^2 + \tau^2).
\end{align}
We refer to model~\eqref{eq-nn-model} as the \emph{normal--normal model}. Under this model, the average treatment effect $\theta^*$ is estimated by the weighted mean
\begin{align}
\label{eq-nn-estimate}
    \hat{\theta} &= \frac{\sum_{k=1}^K w_k(\tau^2)\, \hat{\theta}_k}{\sum_{k=1}^K w_k(\tau^2)},
\end{align}
where $w_k(\tau^2) = 1/(\sigma_k^2 + \tau^2)$. The estimator in \eqref{eq-nn-estimate} is unbiased and coincides with the maximum likelihood estimator under the normal--normal model. A Wald-type test statistic for testing $H_0\!: \theta = \theta_0$ is given by
\begin{align*}
    T_K(\theta) &= 
    \frac{\sum_{k=1}^K w_k(\tau^2) (\hat{\theta}_k - \theta_0)}
         {\sqrt{\sum_{k=1}^K w_k(\tau^2)}}
    \sim N\!\left( (\theta^* - \theta_0) \sqrt{\sum_{k=1}^K w_k(\tau^2)}, 1 \right).
\end{align*}
However, in practice, $\hat{\theta}_k$ may be biased depending on how it is constructed. A representative example is the log odds ratio \citep{bhaumik2012meta}.

\begin{example}[Bias of the log odds ratio]
\label{exam-logor}
{\rm
Suppose $Y_j \sim \mathrm{Binomial}(n_j, p_j)$ with $\mathrm{logit}(p_j) = \mu + j\theta$ for $j = 0, 1$. The log odds ratio with a continuity correction parameter $a \in [0, 1]$ is defined as
\begin{align*}
    \log \mathrm{OR}_a &= 
    \log \frac{Y_{1} + a}{n_1 - Y_{1} + a}
    - \log \frac{Y_{0} + a}{n_0 - Y_{0} + a},
\end{align*}
and its expectation is
\begin{align*}
    E[\log \mathrm{OR}_a] &= \theta 
    + \left(\frac{\frac{1}{2} - a}{n_1}\right)\!\left(e^{\mu + \theta} - e^{-\mu - \theta}\right) 
    - \left(\frac{\frac{1}{2} - a}{n_0}\right)\!\left(e^{\mu} - e^{-\mu}\right) 
    + O(n^{-2}),
\end{align*}
where $n = \min(n_0, n_1)$. When $a = 0$, the bias term $b_k$ is
\begin{align*}
    b_k = 
    \frac{e^{\mu + \theta} - e^{-\mu - \theta}}{2n_1}
    - \frac{e^{\mu} - e^{-\mu}}{2n_0} 
    + O(n^{-2}).
\end{align*}
}
\end{example}

This result shows that a nonlinear transformation of the outcome can induce bias in the expectation of the estimator. Accounting for this bias, model~\eqref{eq-nn-model} can be reformulated as
\begin{align}
\label{eq-biased-nn-model}
    \hat{\theta}_k = \tilde{\theta}_k + b_k, \quad 
    \tilde{\theta}_k \sim N(\theta^*, \sigma_k^2 + \tau^2), \quad 
    b_k = \frac{c_k}{n_k^r} = O(n^{-r}), \qquad r > 0,
\end{align}
where $c_k$ is a constant and $n_k = n a_k$ with $\sum_{k=1}^K a_k = 1$ and $a_k > 0$. In model~\eqref{eq-biased-nn-model}, the bias $b_k$ diminishes as the within-study sample size $n_k$ increases but does not vanish with the number of studies $K$. Hence, even if only one study has a small $n_k$, the overall bias in $\hat{\theta}$ persists and cannot be corrected by increasing $K$.

The following proposition highlights the importance of assumption (A1) for ensuring valid inference under the normal--normal model.

\begin{prop}
\label{lem-false-convergence}
Assume that model~\eqref{eq-biased-nn-model} holds for $\hat{\theta}_k$, $\tilde{\theta}_k$, and $b_k$. Then, the estimator $\hat{\theta}$ defined in \eqref{eq-nn-estimate} satisfies
\begin{align*}
    \hat{\theta} &= \theta^* + \frac{C_K}{n^r} + o_p(1) \quad \text{as } K \to \infty,
\end{align*}
and under the true null hypothesis $H_0\!: \theta = \theta^*$, the test statistic satisfies
\begin{align*}
    P\big(|T_K(\theta^*)| > z_{a}\big) \to 1 \quad \text{as } K \to \infty,
\end{align*}
where 
\[
C_K = 
\frac{\sum_{k=1}^K w_k(\tau^2) c_k / a_k^{-r}}
     {\sum_{k=1}^K w_k(\tau^2)},
\]
and $z_{a}$ denotes the upper $100a\%$ quantile of the standard normal distribution.
\end{prop}

Here, $o_p(1)$ denotes a stochastic term that converges to zero in probability as $K \to \infty$, i.e., for any $\varepsilon > 0$, $P(|o_p(1)| > \varepsilon) \to 0$ as $K \to \infty$. The proof is provided in Appendix~\ref{ap-proof-1}. 

Proposition~\ref{lem-false-convergence} demonstrates that when the study-specific estimators $\hat{\theta}_k$ are biased (i.e., assumption (A1) is violated), the bias cannot be eliminated under the normal--normal model, even as the number of studies increases. Consequently, the Wald-type statistic $T_K(\theta^*)$ diverges under the true null hypothesis, inflating the type~I error rate and producing overly narrow confidence intervals. Conventional random-effects meta-analyses therefore tend to exhibit undercoverage and invalid inference when applied to small or biased studies, consistent with the empirical findings in the simulation studies.

These results provide a theoretical justification for the GLMM-based framework proposed in the main text, which directly models the outcome distribution without relying on the normality assumptions.

\section{Numerical calculation of the likelihood function}

Under the normal--normal model, the likelihood function in equation~(4) of the main text has a closed-form expression. In contrast, when the individual-level outcomes follow a distribution in the exponential family, the likelihood involves an intractable integral that must be approximated numerically. Common numerical integration methods include the Laplace approximation, adaptive Gaussian–Hermite quadrature, and penalized quasi-likelihood \citep{ju2020laplace}. Among these methods, the Laplace approximation and Gaussian–Hermite quadrature offer computational efficiency. However, maintaining integration accuracy across the entire domain $\Omega$ for arbitrary values of $\tau^2$ requires adaptive adjustment of the integration region, which can be computationally intensive if performed separately for each study.

Because analytical integration is infeasible and adaptive quadrature may be computationally demanding, a simpler alternative is Monte Carlo integration:
\begin{align}
\label{eq-Lk}
    \log L_k(\bbe, \tau^2; \bar{y}_k, \x_k, \varphi_k)
    &\propto \log \int_{\Omega} \exp\left( 
        \frac{n_k \left\{ \bar{y}_{k} (\x_k^T \bbe + v_k) - b(\x_k^T \bbe + v_k) \right\}}{a(\varphi_k)} 
    \right) f_V(v_k; \tau^2) \, dv_k \\ \nonumber
    &\approx \log \frac{1}{B} \sum_{b=1}^B \exp\left( 
        \frac{n_k \left\{ \bar{y}_{k} (\x_k^T \bbe + v^{(b)}) - b(\x_k^T \bbe + v^{(b)}) \right\}}{a(\varphi_k)} 
    \right),
\end{align}
where the integration over $v_k$ is independent across studies. A single set of samples $v^{(1)}, \dots, v^{(B)}$ can be drawn independently from $f_V(\cdot; \tau^2)$ and used for all $K$ studies. Monte Carlo integration achieves a convergence rate of $O(B^{-1/2})$, but repeated likelihood evaluations required for profile likelihood estimation can be computationally burdensome. To alleviate this, the convergence rate can be improved to $O(B^{-1} \log B)$ by employing quasi--Monte Carlo integration \citep{sobol2003one}. In quasi--Monte Carlo methods, a low-discrepancy sequence $(u_1, \dots, u_B) \in [0,1]^B$ is generated, and the random-effect samples are obtained as $v^{(b)} = F_V^{-1}(u_b)$, where $F_V^{-1}$ denotes the quantile function of $V$.

\section{Proofs}
\label{app}

Before presenting the proofs, we clarify the asymptotic notation used in this section. Let $\{X_n\}$ be a sequence of random variables and $\{a_n\}$ a sequence of positive constants. We write:
\begin{itemize}
    \item $X_n = o_p(a_n)$ if $X_n / a_n \to 0$ in probability as $n \to \infty$; that is, for any $\varepsilon > 0$, $P(|X_n| > \varepsilon a_n) \to 0$.
    \item $X_n = O_p(a_n)$ if $\{X_n / a_n\}$ is bounded in probability; equivalently, for any $\varepsilon > 0$, there exists $M > 0$ such that $P(|X_n| > M a_n) < \varepsilon$ for sufficiently large $n$.
    \item For deterministic sequences $\{b_n\}$, we write $b_n = O(a_n)$ when $\sup_n |b_n| / a_n < \infty$.
\end{itemize}

Throughout this section, asymptotics with respect to $n$ refer to the common scaling regime
\begin{align*}
n_k = n\,a_k, \qquad a_k > 0, \qquad \sum_{k=1}^K a_k = 1,
\end{align*}
so that each within-study sample size $n_k$ diverges proportionally as $n \to \infty$ while the proportions $\{a_k\}$ remain fixed. Asymptotics with respect to $K$ refer to increasing the number of studies while $\{a_k\}$ and other model components are held fixed. Accordingly, the notation $O_p(K^{-2})$ is defined analogously with respect to $K$.

\subsection{Proof of Proposition \ref{lem-false-convergence}}
\label{ap-proof-1}

\begin{proof}
Under model~\eqref{eq-biased-nn-model}, the overall treatment effect is
\begin{align*}
    \hat{\theta} &= \frac{\sum_{k=1}^K w_k(\tau^2) \hat{\theta}_k}{\sum_{k=1}^K w_k(\tau^2)} 
     = \frac{\sum_{k=1}^K w_k(\tau^2) (\tilde{\theta}_k + b_k)}{\sum_{k=1}^K w_k(\tau^2)} \\
    &= \frac{\sum_{k=1}^K w_k(\tau^2) \tilde{\theta}_k}{\sum_{k=1}^K w_k(\tau^2)} 
       + \frac{\sum_{k=1}^K w_k(\tau^2) b_k}{\sum_{k=1}^K w_k(\tau^2)} \\
    &= \theta^* 
       + \frac{\sum_{k=1}^K w_k(\tau^2) (\tilde{\theta}_k - \theta^*)}{\sum_{k=1}^K w_k(\tau^2)} 
       + \frac{\sum_{k=1}^K w_k(\tau^2) c_k / (n a_k)^r}{\sum_{k=1}^K w_k(\tau^2)} \\
    &= \theta^* + \frac{C_K}{n^r} 
       + \frac{\sum_{k=1}^K w_k(\tau^2) (\tilde{\theta}_k - \theta^*)}{\sum_{k=1}^K w_k(\tau^2)} 
       = \theta^* + \frac{C_K}{n^r} + o_p(1).
\end{align*}

From this, we obtain
\begin{align*}
    T_K &= \left(\sum_{k=1}^K w_k(\tau^2) \right)^{1/2} (\hat{\theta} - \theta^*) \\
    &= \left(\sum_{k=1}^K w_k(\tau^2) \right)^{1/2} \frac{C_K}{n^r} 
       + \frac{\sum_{k=1}^K w_k(\tau^2) (\tilde{\theta}_k - \theta^*)}
              {\sqrt{\sum_{k=1}^K w_k(\tau^2)}} 
       \xrightarrow{p} \pm \infty,
\end{align*}
as $K \to \infty$. Since $T_K \xrightarrow{p} \pm \infty$, for any fixed constant $c > 0$,
\[
    P(|T_K| > c \mid \theta^*) \to 1 \quad \text{as } K \to \infty.
\]
In particular, for $c = z_{\alpha/2}$,
\[
    P(|T_K| > z_{\alpha/2} \mid \theta^*) \to 1 \quad \text{as } K \to \infty.
\]
\end{proof}

\subsection{Proof of Lemma 1}
\label{ap-proof-2}

\begin{proof}
The likelihood function conditional on a fixed $V_k = v_k$ can be written as
\begin{align*}
    L_k(\bbe; y_{k1}, \dots, y_{kn_k}, \x_k, v_k, \varphi_k) 
    &= \prod_{i=1}^{n_k} f_Y(y_{ki}; \bbe, v_k) \\
    &= \prod_{i=1}^{n_k} 
       \exp\!\left( 
           \frac{y_{ki}(\x_k^T \bbe + v_k) - b(\x_k^T \bbe + v_k)}{a(\varphi_k)} 
           + c(y_{ki}, \varphi_k) 
       \right) \\
    &= \left( \prod_{i=1}^{n_k} e^{c(y_{ki}, \varphi_k)} \right)
       \exp\!\left( 
           \sum_{i=1}^{n_k} 
           \frac{y_{ki}(\x_k^T \bbe + v_k) - b(\x_k^T \bbe + v_k)}{a(\varphi_k)} 
       \right) \\
    &= \left( \prod_{i=1}^{n_k} e^{c(y_{ki}, \varphi_k)} \right)
       \exp\!\left( 
           \frac{n_k[\bar{y}_{k}(\x_k^T \bbe + v_k) - b(\x_k^T \bbe + v_k)]}{a(\varphi_k)} 
       \right),
\end{align*}
where $\bar{y}_k = n_k^{-1} \sum_{i=1}^{n_k} y_{ki}$ is the sample mean. Integrating over $v_k$, the marginal likelihood becomes
\begin{align*}
    L_k(\bbe, \tau^2) &= E_{V}\![L_k(\bbe; y_{k1}, \dots, y_{kn_k}, \x_k, v_k, \varphi_k)] \\
    &= \int_{-\infty}^{\infty} L_k(\bbe; y_{k1}, \dots, y_{kn_k}, \x_k, v_k, \varphi_k)
       f_V(v_k; \tau^2)\, dv_k \\
    &= \left( \prod_{i=1}^{n_k} e^{c(y_{ki}, \varphi_k)} \right)
       \int_{-\infty}^{\infty} 
       \exp\!\left( 
           \frac{n_k[\bar{y}_{k}(\x_k^T \bbe + v_k) - b(\x_k^T \bbe + v_k)]}{a(\varphi_k)} 
       \right)
       f_V(v_k; \tau^2)\, dv_k.
\end{align*}
Since the likelihood depends on $\{y_{ki}\}$ only through $\bar{y}_k$, it follows from the factorization theorem that $(\bar{y}_1, \dots, \bar{y}_K)$ is a sufficient statistic for $(\bbe, \tau^2)$.
\end{proof}

\subsection{Proof of Theorem 1}
\label{ap-proof-3}

\begin{proof}
For fixed $\beta_{\ell}$,
\begin{align*}
    |T_{SBC}(\beta_{\ell}) - T_{BC}(\beta_{\ell})| 
    &= \left| \frac{T(\beta_{\ell})}{1 + 2C_{SBC}(\beta_{\ell})} - \frac{T(\beta_{\ell})}{1 + 2C_{BC}(\beta_{\ell})} \right| \\
    &= \left| \frac{T(\beta_{\ell})}{1 + 2C_{BC}(\beta_{\ell})}\left( \frac{1 + 2C_{BC}(\beta_{\ell})}{1 + 2C_{SBC}(\beta_{\ell})} - 1 \right) \right| \\
    &\le \left| \frac{2T(\beta_{\ell})}{(1 + 2C_{BC})(1 + 2C_{SBC})} \right| \left| C_{BC}(\beta_{\ell}) - C_{SBC}(\beta_{\ell}) \right| \\
    &= M_1 | C_{BC}(\beta_{\ell}) - C_{SBC}(\beta_{\ell}) |,
\end{align*}
where 
\begin{align*}
    M_1 = \left| \frac{2T(\beta_{\ell})}{(1 + 2C_{BC})(1 + 2C_{SBC})} \right| \le 2|T(\beta_{\ell})|.
\end{align*}
By \citet{wilks1938large}, $T(\beta_{\ell}) = \chi_1^2 + O_p(K^{-1/2})$, hence $M_1 = O_p(1)$.

Next, $|C_{BC}(\beta_{\ell}) - C_{SBC}(\beta_{\ell})| = O(n^{-1/2})$. Using \citet{lawley1956general},
\begin{align*}
    C_{BC}(\beta_{\ell}) &= \frac{1}{2K} \{l_2^{-2}(\tfrac{1}{4}l_4 - l_{31} + l_{22}) - l_2^{-3}(\tfrac{5}{12}l_3^2 - 2l_3l_{21} + 2l_{21}^2)\},
\end{align*}
where $l_r = E[\partial^r l / \partial \beta_{\ell}^r]$, $l_{rs} = \partial^s l_r / \partial \beta_{\ell}^s$, and $l(\beta_{\ell}) = \sum_{k=1}^K \log L_k(\hat{\bbe}(\beta_{\ell}), \tilde{\tau}^2(\beta_{\ell}))$.

Because $Y$ follows an exponential-family distribution with $E[|Y|^s] < \infty$ for $s = 1, 2, 3$, Theorem~2 of \citet{bhattacharya1978validity} implies that as $n_k \to \infty$,
\begin{align*}
    \log L_k(\bbe, \tau^2)
    = \log \phi(g(\bar{y}_k); \x_k^T \bbe, \sigma_k^2 + \tau^2)
    + O(n_k^{-1/2}),
\end{align*}
where $\phi$ denotes the normal density with mean $\x_k^T \bbe$ and variance $\sigma_k^2 + \tau^2$. 
Let
\begin{align*}
    \hat{l}(\beta_\ell)
    = \sum_{k=1}^K
    \log \phi\bigl(
        g(\bar{y}_k);
        \x_k^T \hat{\bbe}(\beta_\ell),
        \sigma_k^2 + \tilde{\tau}^2(\beta_\ell)
    \bigr).
\end{align*}
Then we have $l(\beta_{\ell}) = \hat{l}(\beta_{\ell}) + O(n^{-1/2})$, and the derivatives satisfy $l_r^m = \hat{l}_r^m + O(n^{-1/2})$. 
Substituting these expressions yields $C_{BC}(\beta_{\ell}) = C_{SBC}(\beta_{\ell}) + O(n^{-1/2})$.

Hence $|C_{BC} - C_{SBC}| = O(n^{-1/2})$ and $M_1 = O_p(1)$, so $|T_{SBC} - T_{BC}| = O(n^{-1/2}) \cdot O_p(1) = O_p(n^{-1/2})$. Since $T_{BC}(\beta_{\ell}) = \chi_1^2 + O(K^{-2})$,
\[
    T_{SBC}(\beta_{\ell}) = \chi_1^2 + O_p(n^{-1/2} + K^{-2}).
\]
\end{proof}

\section{Simulation details for each individual distribution}
\label{app-sim}

This section describes the parameter settings for the normal, binomial, Poisson, and gamma distributions used in the simulation study.

\subsection{Normal distribution $N(\mu_k, \sigma^2)$}
The parameters in the density function of $Y_{ki}$ given in (3) in the main text are specified as follows:
\begin{align*}
    \theta_k &= \mu_k, \quad
    a(\varphi_k) = \sigma^2, \quad
    b(\theta_k) = \frac{\theta_k^2}{2}, \quad
    c(y_{ki}, \varphi_k) = -\frac{y_{ki}^2}{2\sigma^2} - \frac{1}{2}\log(2\pi\sigma^2).
\end{align*}
With these specifications, the density function becomes
\begin{align*}
    f_Y(y_{ki})
    = \exp\!\left\{ -\frac{(y_{ki} - \mu_k)^2}{2\sigma^2} - \frac{1}{2} \log(2\pi\sigma^2) \right\}.
\end{align*}
This corresponds to a model in which the outcome follows a normal distribution with an identity link function.

In the simulation, the within-study variance was set to $\sigma^2 = 50 / n_k$, corresponding to an individual-level variance of 50 and aggregation over $n_k$ subjects within each study. Accordingly, the study-specific mean $\bar{y}_k$ follows a normal distribution:
\begin{align*}
    \bar{y}_k \sim N(\mu_k, 50 / n_k).
\end{align*}
This configuration serves as a baseline for comparison with the non-normal outcome settings in the simulation study.

\subsection{Binomial distribution $Bin(n_k, p_k)$}
The parameters in the density function of $Y_{ki}$ given in (3) in the main text are specified as follows:
\begin{align*}
    \theta_k &= \log \frac{p_k}{1 - p_k}, \quad 
    a(\varphi_k) = 1, \quad 
    b(\theta_k) = n_k \log(1 + e^{\theta_k}), \quad
    c(y_{ki}, \varphi_k) = - \log \left( \begin{array}{c}
        n_k \\
        y_{ki} 
    \end{array} \right).
\end{align*}
With these specifications, the density function is given by
\begin{align*}
    f_Y(y_{ki}) 
    = \exp\!\left\{
        y_{ki}\theta_k 
        - n_k \log(1 + e^{\theta_k}) 
        + \log \left( \begin{array}{c}
            n_k \\
            y_{ki}
        \end{array} \right)
      \right\}.
\end{align*}
This corresponds to a model in which the outcome follows a binomial distribution with the canonical logit link function. Because the dispersion parameter $\varphi_k$ does not appear in the binomial distribution, no estimation of nuisance parameters is required.

\subsection{Poisson distribution $Poisson(\mu_k)$}
The parameters in the density function of $Y_{ki}$ given in (3) in the main text are specified as follows:
\begin{align*}
    \theta_k &= \log \mu_k = \log t_{ki} + \x_k^T \bbe + v_k, \quad
    a(\varphi_k) = 1, \quad
    b(\theta_k) = e^{\theta_k} = \mu_k = t_{ki} e^{\x_k^T \bbe + v_k}, \quad
    c(y_{ki}, \varphi_k) = -\log(y_{ki}!).
\end{align*}
With these specifications, the density function is given by
\begin{align*}
    f_Y(y_{ki}) 
    &= \exp\!\left\{ y_{ki} \theta_k - e^{\theta_k} - \log(y_{ki}!) \right\} \\
    &= \exp\!\left\{ 
        y_{ki}(\log t_{ki} + \x_k^T \bbe + v_k)
        - t_{ki} e^{\x_k^T \bbe + v_k}
        - \log(y_{ki}!)
      \right\} \\
    &= \exp\!\left\{
        y_{ki}(\x_k^T \bbe + v_k)
        - t_{ki} e^{\x_k^T \bbe + v_k}
        + y_{ki}\log t_{ki}
        - \log(y_{ki}!)
      \right\}.
\end{align*}
This formulation corresponds to a model in which the outcome follows a Poisson distribution with the canonical log link function. In the simulation, the observed time $t_{ki}$ was fixed at 1 to focus exclusively on the estimation of $\bbe$, simplifying the model without loss of generality. Because the dispersion parameter $\varphi_k$ does not appear in the Poisson distribution, no estimation of nuisance parameters is required.

\subsection{Gamma distribution $Gamma(a, b)$}
Let $a > 0$ denote the shape parameter and $b > 0$ the scale parameter. The distribution can be reparameterized as $a = 1/\varphi_k$ and $b = \varphi_k / \mu_k$. The parameters in the density function of $Y_{ki}$ given in (3) in the main text are specified as follows:
\begin{align*}
    \theta_k &= -\frac{1}{\mu_k}, \quad
    a(\varphi_k) = \varphi_k, \quad
    b(\theta_k) = -\log(\theta_k), \quad
    c(y_{ki}, \varphi_k) = \left(\frac{1}{\varphi_k} - 1\right)\log(y_{ki})
    - \frac{1}{\varphi_k}\log(\varphi_k)
    - \log \Gamma\!\left(\frac{1}{\varphi_k}\right).
\end{align*}
With these specifications, the density function is expressed as
\begin{align*}
    f_Y(y_{ki})
    = \exp\!\left\{
        \frac{y_{ki}\theta_k + \log(\theta_k)}{\varphi_k}
        + \left(\frac{1}{\varphi_k} - 1\right)\log(y_{ki})
        - \frac{1}{\varphi_k}\log(\varphi_k)
        - \log \Gamma\!\left(\frac{1}{\varphi_k}\right)
      \right\}.
\end{align*}
This formulation corresponds to a model in which the outcome follows a gamma distribution with a log link function. While the binomial and Poisson models employ the canonical link, the canonical link for the gamma distribution is $g(\mu_k) = -1/\mu_k$, which requires $\mu_k > 0$ and can lead to numerical instability. Therefore, the log link function was adopted in this study.

In the simulation, the dispersion parameter $\varphi_k$ was set as
\begin{align*}
    \varphi_k = \frac{1}{3} \left(1 + 4 \frac{k - 1}{K}\right), \quad k = 1, \dots, K.
\end{align*}
In practice, $\varphi_k$ is typically unknown and can be estimated from the sample mean $\bar{y}_k$ and sample variance $\hat{\sigma}_k^2$ in each study as $\hat{\varphi}_k = \hat{\sigma}_k^2 / \bar{y}_k^2$, which is then substituted into $f_Y(\cdot)$. For the normal, binomial, and Poisson models, the dispersion parameter $\varphi_k$ is fixed (equal to the variance or unity), whereas for the gamma model, it varies across studies and is estimated from the study-level mean and variance.

\end{document}